\documentclass[reprint,amsmath,amssymb,aps,superscriptaddress, showpacs, eprint, prl]{revtex4-2}
\usepackage[utf8]{inputenc}

\usepackage[english]{babel}

\usepackage{amsmath,amssymb,ascmac,fancybox, multirow}
\usepackage[english]{babel} 
\usepackage{color}
\usepackage{graphicx}
\usepackage{url}
\usepackage{siunitx}
\usepackage{bm}
\usepackage{physics}
\usepackage{mathtools}
\usepackage{hyperref}
\usepackage[T1]{fontenc}
\usepackage{tensor}
\usepackage{ulem}
\graphicspath{figures/}

\newcommand{\ani}[1]{{#1}}
\newcommand{\cre}[1]{{#1}^\dagger}

\renewcommand{\vec}{\vb*}

\newcommand{\isoS}{\bar{S}}

\def \be{\begin{equation}}
\def \ee{\end{equation}}
\def \bea{\begin{eqnarray}}
\def \eea{\end{eqnarray}}

\begin{document}

\title{Geometric bound on structure factor}%

\author{Yugo Onishi}
\affiliation{Department of Physics, Massachusetts Institute of Technology, Cambridge, MA 02139, USA}
\author{Alexander Avdoshkin}
\affiliation{Department of Physics, Massachusetts Institute of Technology, Cambridge, MA 02139, USA}
\author{Liang Fu}
\affiliation{Department of Physics, Massachusetts Institute of Technology, Cambridge, MA 02139, USA}
\date{\today}

\begin{abstract}
We show that a quadratic form of quantum geometric tensor in $k$-space sets a bound on the $q^4$ term in the static structure factor $S(q)$ at small $\vec{q}$. Bands that saturate this bound satisfy a condition similar to Laplace's equation, leading us to refer to them as \textit{harmonic bands}.  We provide examples of harmonic bands in one- and two-dimensional systems, including (higher) Landau levels.  
The geometric bound further leads to a topological bound on the $q^4$ term, which is saturated only when the band geometry satisfies the trace condition and, additionally, the quantum geometric tensor is uniform in $k$-space. We speculate that these bounds taken together provide
a useful guide for identifying Chern bands that favor (Abelian or non-Abelian) fractional Chern insulators.
\end{abstract}

\maketitle
\textit{Introduction. ---} The static structure factor is an important physical observable that characterizes density correlations in ground states of electronic systems. While the Bragg peaks in the structure factor have been used to identify periodic crystal structures and charge density wave orders, the structure factor away from reciprocal lattice vectors also encodes rich information about the electronic structure and physical responses of the system. For example, the structure factor is related to the density responses through the fluctuation-dissipation theorem~\cite{callen_irreversibility_1951}. In interacting systems, it determines the interaction energy of the ground state, and also offers a way to estimate the excitation energy, as shown by Feynman for liquid Helium systems~\cite{feynman_atomic_1954}. A related approach based on the projected structure factor was developed to study the collective modes in fractional quantum Hall systems~\cite{girvin_magneto-roton_1986}. 

Recently, an exact general relation between the structure factor and the many-body quantum geometry was established~\cite{onishi2024h}, enabling direct measurement of the ground state quantum geometry via X-ray scattering and electron loss spectroscopy~\cite{balut_quantum_2024}. 
Specifically, for insulating states in reduced dimensions or with short-range interaction, the static structure factor at leading order in the wavevector $q$ is directly related to the many-body quantum metric. 
The many-body quantum metric is also closely related to polarization fluctuation~\cite{souza_polarization_2000}, localization of electron wavefunction~\cite{resta_electron_1999}, the spread of Wannier functions~\cite{marzari_maximally_1997}, and the Bloch band geometry in noninteracting band insulators. Furthermore, this geometric perspective naturally leads to universal bounds on physical observables in topological phases~\cite{onishi_topological_2024, onishi2024h, batra_bound_2024, yu_universal_2024}. %

In this work, we explore the relation between quantum geometry and the static structure factor $S(\vec{q})$ of band insulators beyond the leading order in $q$. 
Specifically, we show that the isotropic part of $S(\vec{q})$ at fourth order in $q$ (denoted as $S_4$) has a universal bound determined by the quadratic form of the quantum geometric tensor.
This establishes a connection between the structure factor and fluctuations of quantum geometric tensor in $k$-space.
Interestingly, we find that this geometric bound is saturated when the band geometry satisfies a certain ``harmonic'' condition. Examples of harmonic bands in one- and two-dimensional systems, including higher Landau levels, are provided.

This geometric bound allows us to derive a topological bound on $S_4$. 
This bound is saturated only when the band geometry satisfies the trace condition~\cite{roy_band_2014, claassen_position-momentum_2015, ledwith_fractional_2020, mera_kahler_2021, wang_exact_2021, ledwith_family_2022, ledwith_vortexability_2023, liu_theory_2024} \textit{and} is uniform in $k$-space.
Our geometric and topological bounds on $S_4$ offer a useful guide for identifying Chern bands that are promising for hosting fractional Chern insulators, including non-Abelian ones.

In this work, we define the static structure factor as the following \textit{equal-time} density-density correlation function:
\begin{align}
    S(\vec{q}) &\equiv \frac{1}{N_e}\expval{{n}_{\vec{q}}{n}_{-\vec{q}}}
\end{align}
where ${n}_{\vec{q}}=\int\dd{r}e^{-i\vec{q}\vdot\vec{r}}n(\vec{r})$ is the number density operator at wavevector $\vec{q}$, and $N_e$ is the total number of electrons. In tight-binding models, $n_{\vec{q}}$ is defined as $n_{\vec{q}}=\sum_{i}e^{-i\vec{q}\vdot\vec{r}_i}$ with $\vec{r}_i$ the position of site $i$. With this convention, $S(\vec{q}=0)=N_e$. Note that this normalization differs from that used in Refs.~\cite{onishi_topological_2024, onishi2024h}.

\textit{Atomic insulators. ---} 
To gain intuition about the structure factor of band insulators at small $q$, let us start with a simple example of an array of atoms forming a $d$-dimensional cubic 
lattice. When the lattice constant $a$ is large enough, we can neglect the hopping between the atoms and the system is an atomic insulator. Its energy eigenstates are given by the localized orbitals of individual atoms, %
denoted as $\phi_n(\vec{r}-\vec{R})$, with $\vec{R}$ the position of an atom and $n$ the label for each atomic orbital. When each atom has one electron occupying $n=0$ orbital, the structure factor at $\vec{q}$ away from the reciprocal lattice vectors is given by (see SM for the derivation)
\begin{align}
    S(\vec{q}) &= 1-\abs{\expval{e^{-i\vec{q}\vdot\vec{r}}}_0}^2
    \label{Sq-AI}
\end{align}
where $\expval{\dots}_0 = \int\dd{r} (\dots) \abs{\phi_0(\vec{r})}^2$ is the expectation value in $n=0$ orbital. 
The quantity $\expval{e^{-i\vec{q}\vdot\vec{r}}}_0$ %
contains all the moments of $\vec{r}$ for the probability distribution $p(\vec{r})=\abs{\phi_0(\vec{r})}^2$. Thus, the structure factor completely captures electron density distribution within an atom. 

At small $\vec{q}$, we expand $S(\vec{q})$ in $\vec{q}$ and obtain 
\begin{align}
    S(\vec{q}) 
    &= \expval{(\vec{q}\vdot\vec{r})^2}_0 - \frac{2}{4!}\qty(\expval{(\vec{q}\vdot\vec{r})^4}_0 + 3\expval{(\vec{q}\vdot\vec{r})^2}_0^2) + \dots \nonumber \\
    &= K_{\alpha\beta} \frac{\bar{q}_{\alpha}\bar{q}_{\beta}}{2\pi}  + \dots, %
\end{align}
where $\bar{q}_{\alpha}=q_{\alpha}a$ is the dimensionless wavevector.
Here, we have chosen the origin such that $\expval{\vec{r}}_0=0$, and $K$ is a rank-2 tensor given by
\begin{align}
    K_{\alpha\beta} &= 2\pi\frac{\expval{r_{\alpha}r_{\beta}}_0}{a^2}, \label{eq:K_atom}
\end{align}
which represents the position fluctuation of an electron within the atom. 

Although $q^4$ term is more complicated, its isotropic component is fairly simple. We define the isotropic component of $S(\vec{q})$ as $\isoS(q)=\int_{\abs{\vec{q}}=q}\dd{\Omega} S(\vec{q})/A_{d}$, where $d$ is the spatial dimension, $\int\dd\Omega$ is integration over solid angle, and $A_{d}$ is the total solid angle in $d$-dimensions. We expand $\isoS(q)$ in $q$ as 
\begin{align}
    \isoS(q) &= \frac{S_2}{2!}\frac{\bar{q}^2}{2\pi} - \frac{S_4}{4!}\qty(\frac{\bar{q}^2}{2\pi})^2 + \dots, 
\end{align}
where $S_2$ and $S_4$ are given by 
\begin{align}
    S_2 &= \frac{4\pi }{da^{2}}\expval{r^2}_0 = \frac{2}{d}\tr K,\\
    S_4 &= \frac{12}{d(d+2)} \qty((\tr K)^2 + \tr (K^2)) \nonumber \\
    & \quad + \frac{6 (2\pi)^2}{d(d+2) a^4} \expval{(r^2-\expval{r^2})^2}_0.  \label{eq:S4_atom}
\end{align}
Here and hereafter, $\tr$ represents the sum over the spatial indices $x, y, \dots$. 
In probability theory, the fourth moment, such as $\expval{x^4}$, is referred to as kurtosis (up to appropriate normalization), thus we shall call $S_4$ \textit{kurtosis} from now on.
Since the last term in $S_4$ is semi-positive, we see a lower bound on $S_4$ determined solely by
the quadratic coefficients of structure factor, $K$: 
\begin{align}
    S_4 \ge \frac{12}{d(d+2)} \qty((\tr K)^2 + \tr (K^2)). \label{eq:S4_bound_atom}
\end{align}
Notably, the bound on $S_4$ is set by the square of $K$. Since $K$ is related to quantum metric~\cite{onishi_quantum_2024}, this observation suggests a relation between $S_4$ and the quadratic form of quantum geometry.

In the rest of this work, we will present a generalization of Eq.~\eqref{eq:S4_bound_atom} that applies to all band insulators. That is, the isotropic part of $S(q)$ at $q^4$ order has a lower bound given by the quadratic form of quantum geometry in $k$-space, involving both the quantum metric and Berry curvature.

\textit{General theory. ---} Consider a $d$-dimensional non-interacting band insulator. The static structure factor at finite small $\vec{q}$ is given by~\cite{onishi2024h} 
\begin{align}
    S(\vec{q}) %
    &= \frac{1}{\nu}\int[dk]\Tr[P(\vec{k})(P(\vec{k})-P(\vec{k}+\vec{q}))] 
    \label{Sq}
\end{align}
where $P(\vec{k})=\sum_{n=1}^{\nu}\ketbra{u_n(\vec{k})}{u_n(\vec{k})}$ is the projector onto the occupied bands at wavevector $\vec{k}$ with the {\it cell-periodic} Bloch wavefunction of $n$-th band $\ket{u_n(\vec{k})}$ and the number of occupied bands $\nu$. $\int[dk] = \int_{\rm BZ}V_{\rm uc} d^dk/(2\pi)^d$ is the integral over the Brillouin zone normalized by the unit cell volume $V_{\rm uc}$. We can confirm $S(\vec{q}\to\infty)=1$ since $\lim_{\vec{q}\to \infty} \Tr[P(\vec{k})P(\vec{k}+\vec{q})]=0$ and $\Tr[P(\vec{k})^2]=\Tr[P(\vec{k})]=\nu$. 
For simplicity, we first consider cases with $\nu=1$ so that the electron density $n$ is $n=1/V_{\rm uc}$, before addressing the general cases. 

As shown by Eq.~\eqref{Sq}, the static structure factor of band insulators is determined solely by the Bloch wavefunction of occupied bands, suggesting a connection between the structure factor and quantum geometry. The connection becomes clear by expanding $S(\vec{q})$ in powers of $q$:  
\begin{align}
    S(\vec{q}) &= K_{\alpha\beta} \frac{\bar{q}_{\alpha}\bar{q}_{\beta}}{2\pi} + \dots,
\end{align}
where $\bar{q}_{\alpha}\equiv q_{\alpha}/n^{1/d}=q_{\alpha}a$ is the dimensionless wavevector with $a\equiv V_{\rm uc}^{1/d}$.
The quadratic coefficient $K$, recently termed the quantum weight, 
is directly given by the quantum metric of the occupied bands~\cite{onishi_topological_2024}
\begin{align}
    K_{\alpha\beta} &= 2\pi \int  [\dd{k}] g_{\alpha\beta},
\end{align}
where the quantum metric $g = \Re Q$ is the real part of the quantum geometric tensor defined as 
\begin{align}
    Q_{\alpha\beta} = \frac{1}{a^2}\mel{\partial_{\alpha}u}{(1-P)}{\partial_{\beta}u} = \frac{1}{a^2}\ip{D_{\alpha}u}{D_{\beta}u}. \label{eq:QGT}
\end{align}
Here, $D_{\alpha}=\partial_\alpha + i A_{\alpha}$ is the covariant derivative with $A_{\alpha}=\mel{u}{i\partial_{\alpha}}{u}$ the (Abelian) Berry connection. We define $Q$ with normalization $1/a^2$ so that quantum geometric quantities are dimensionless. 

In the special case of atomic insulators, the cell-periodic Bloch wavefunction $\ket{u_n(\vec{k})}= \sum_{\vec{R}} \exp(i \vec{k} \cdot (\vec{x}-\vec{R}) ) \ket{\phi_n(\vec{R})}$ is a linear combination of atomic orbitals. Then, the structure factor reduces to the expression Eq.~\eqref{Sq-AI} and the quantum metric tensor $g_{\alpha \beta}(\vec{k})$ becomes a constant given by the expectation value of quadrupole moment $r_\alpha r_\beta$ within an atom as shown in 
Eq.~\eqref{eq:K_atom}~\cite{onishi_fundamental_2024}.

Our focus in this work is on higher-order terms in the small-$q$ expansion of $S(\vec{q})$. In particular, we consider the $q$-expansion of its isotropic part:
\begin{align}
    \bar{S}(q) &\equiv \int_{\vec{q}=q}\frac{\dd{\Omega}}{A_d} S(\vec{q}) 
    = \frac{S_2}{2!}\frac{\bar{q}^2}{2\pi} - \frac{S_4}{4!}\qty(\frac{\bar{q}^2}{2\pi})^2 + \dots, \label{eq:isoS_expansion}
\end{align}
where $S_2 = \tr K$ and $S_4$ is the next leading order term of our interest. We can express $S_4$ in a simple form: 
\begin{align}
    S_4 &= \frac{3(2\pi)^2}{d(d+2)a^4}\int[\dd{k}] \Tr[(\nabla^2 P)^2]. \label{eq:S4_2dim}
\end{align}
We further rewrite $S_4$ with the Bloch wavefunctions as
\begin{align}
    S_4 
    &= \frac{12(2\pi)^2}{d(d+2)}\int[\dd{k}] \qty((\tr Q)^2 + \tr (Q^2)) \nonumber \\
    &\quad + \frac{6(2\pi)^2}{d(d+2)a^4}\int[\dd{k}] \mel{D^2 u}{(1-P)}{D^2 u}. \label{eq:S4_general}
\end{align}
Eq.~\eqref{eq:S4_general} is the generalization of Eq.~\eqref{eq:S4_atom} to arbitrary band insulators with one occupied band.
It reduces to Eq.~\eqref{eq:S4_atom} for atomic insulators since the quantum metric is uniform in $\vec{k}$-space and the Berry curvature vanishes. 

Noting that the last term is semi-positive as $\mel{D^2 u}{(1-P)}{D^2 u}=\norm{(1-P)\ket{D^2 u}}^2 \ge 0$ and using $\tr Q = \tr g$ and $\tr (Q^2) = \tr(g^2) - \tr(\Omega^2)/4$, we obtain a bound for $S_4$ in terms of the quantum geometric tensor:
\begin{align}
   S_4  &\ge \frac{12(2\pi)^2}{d(d+2)}\int[\dd{k}] \qty((\tr Q)^2 + \tr (Q^2)) \\
   &=\frac{12(2\pi)^2}{d(d+2)}\int[\dd{k}] \qty((\tr g)^2 + \tr(g^2) - \frac{\tr(\Omega^2)}{4}). \label{eq:geometric_bound}
\end{align}
Eq.~\eqref{eq:geometric_bound} is the first main result of this work. Importantly, the bound on $S_4$ is set by the quadratic form of the quantum geometric tensor $Q$ in $k$-space. This should be contrasted with $S_2$, which involves only the integral of quantum metric.
We also note that the general geometric bound~\eqref{eq:geometric_bound} involves the Berry curvature, which is absent in the atomic insulator example discussed above.

The geometric bound~\eqref{eq:geometric_bound} is saturated when the last term in Eq.~\eqref{eq:S4_general} vanishes, or equivalently,
\begin{align}
    (1-P)\ket{D^2 u} = 0. \label{eq:saturate_condition}
\end{align}
When the system has time reversal and inversion symmetry, the Berry connection $\vec{A}$ vanishes and $D_{\alpha} = \partial_{\alpha}$. In this case, the saturation condition~\eqref{eq:saturate_condition} becomes
    $(1-P)\ket{\nabla^2 u} = 0$.
This condition can be viewed as a generalized form of Laplace's equation with the factor $1-P$. Therefore, we shall call a band satisfying Eq.~\eqref{eq:saturate_condition} a \textit{harmonic band}, analogous to harmonic functions that satisfy the regular Laplace's equation. When the filled band is harmonic, the geometric bound~\eqref{eq:geometric_bound} is saturated.

Finally, we note that the last term in Eq. \eqref{eq:S4_general}, $\mel{D^2 u}{(1-P)}{D^2 u}$, cannot be written in terms of the quantum metric and Berry curvature. Instead, it represents a new kind of geometry contained in the structure factor. A detailed study will be reported elsewhere~\cite{onishi2025}. %

\textit{Two-band model example. ---} 
To illustrate our geometric bound, we consider a two-band model with the lower band fully occupied. The Hamiltonian can be written as
$H = -E(\vec{k}) \hat{\vec{n}}_{\vec{k}}\vdot \vec{\sigma}$, %
where $\vec{\sigma}=(\sigma_x, \sigma_y, \sigma_z)$ are the Pauli matrices, $E(\vec{k})>0$ determines the band dispersion, and $\hat{\vec{n}}_{\vec{k}}$ is a unit vector. The projector for the lower band is $P(\vec{k})=(1+\hat{\vec{n}}_{\vec{k}}\vdot\vec{\sigma})/2$, and thus the structure factor is $S(\vec{q})=(1/2)\int[\dd{k}] \hat{\vec{n}}_{\vec{k}}\vdot(\hat{\vec{n}}_{\vec{k}}-\hat{\vec{n}}_{\vec{k}+\vec{q}})$, and in particular $S_4$ is given by
\begin{align}
    S_4 &= \frac{3(2\pi)^2}{2d(d+2)a^4}\int[\dd{k}] (\nabla^2\hat{\vec{n}})^2 \label{eq:L_twoband}
\end{align}
On the other hand, the quantum metric $g_{\alpha\beta}$ and the Berry curvature $\Omega_{\alpha\beta}$ are given by $g_{\alpha\beta} = (4a^2)^{-1}(\partial_{\alpha}\hat{\vec{n}}_{\vec{k}})\vdot(\partial_{\beta}\hat{\vec{n}}_{\vec{k}})$ and $\Omega_{\alpha\beta} = -(2a^2)^{-1}\hat{\vec{n}}\vdot(\partial_\alpha\hat{\vec{n}} \cross \partial_\beta\hat{\vec{n}})$, and thus the geometric bound~\eqref{eq:geometric_bound} is given by 
\begin{align}
    S_4 \ge \frac{6\pi^2}{d(d+2)a^4} \int[\dd k]\qty(\vec{n}\vdot\nabla^2\vec{n})^2.
\end{align}
We can verify the bound with an inequality $(\nabla^2\hat{\vec{n}})^2\ge (\hat{\vec{n}}\vdot\nabla^2\hat{\vec{n}})^2$. The bound is saturated when $\nabla^2 \hat{\vec{n}} \parallel \hat{\vec{n}}$.

One simple example saturating the geometric bound is a one-dimensional two-band system where $\hat{\vec{n}}=(\cos\theta, \sin\theta, 0)$ with $\theta=m k_x a$, the lattice constant $a$, and an integer $m$. In this case, the Berry curvature vanishes, and the quantum metric is uniform over the Brillouin zone, given by $g_{xx}=m^2/4$. Then $S_4 = 2\pi^2 m^4$, saturating the geometric bound. 
This two-band model with $m=1$ is realized in a one-dimensional tight-binding model with two orbitals per site, labeled $\alpha=a, b$. Denoting the annihilation operator for orbital $\alpha$ at $n$-th site by $\ani{c}_{n,\alpha}$, the Hamiltonian in real space is given by  
   $H = \sum_{n} \qty(t \cre{c}_{n+1,a}\ani{c}_{n, b} + \text{h.c.})$. %

It is worth noting that the harmonic condition in one-dimensional two-band systems has a geometric interpretation.
The harmonic condition in one dimension reduces to $(\partial_x^2 \hat{\vec{n}})_{\perp}=0$, with $(\dots)_{\perp}$ denoting the component perpendicular to $\hat{\vec{n}}$. This is equivalent to that a curve specified by $\hat{\vec{n}}$ is geodesic on the sphere.

\textit{Generalization to multiband cases. ---}
We now generalize the above discussion to systems with multiple occupied bands. To this end, we introduce the non-Abelian quantum geometric tensor as~\cite{ma_abelian_2010}  
\begin{align}
    \mathcal{Q}_{\alpha\beta} &= \frac{1}{a^2}(\partial_{\alpha} P)(1-P)(\partial_{\beta} P), \label{eq:non-abelian_QGT}
\end{align}
where $P$ is the projection operator for all occupied bands. $\mathcal{Q}_{\alpha\beta}$ is a matrix whose $(i,j)$ component for occupied bands $i,j$ is given by $\mathcal{Q}^{ij}_{\alpha\beta} = a^{-2}\mel{\partial_{\alpha}u_i}{(1-P)}{\partial_{\beta}u_j}$.
Note that the trace of $\mathcal{Q}_{\alpha\beta}$ over the band indices, $Q_{\alpha\beta}\equiv\Tr\mathcal{Q}_{\alpha\beta}$, yields the Abelian quantum geometric tensor for the Slater determinant state composed of all occupied states at $\vec{k}$, $\ket{\Psi_{\vec{k}}}=|u_{\vec{k}1}\dots u_{\vec{k}\nu}|$~\cite{onishi_fundamental_2024}.

With the non-Abelian quantum geometric tensor~\eqref{eq:non-abelian_QGT}, we can find the geometric bound on $S_4$ for general band insulators (see Supplemental Material for the derivation)
\begin{align}
    S_4 &\ge \frac{12(2\pi)^2\nu^{4/d-1}}{d(d+2)}\int[\dd{k}] \Tr[\mathcal{Q}_{\alpha\alpha}\mathcal{Q}_{\beta\beta}+\mathcal{Q}_{\alpha\beta}\mathcal{Q}_{\beta\alpha}], \label{eq:multiband_bound}
\end{align}
where $\nu$ is the number of occupied bands and the repeated spatial indices $\alpha,\beta=x,y$ are summed over. Note that $\bar{q}=qn^{-1/d}$ used to define $S_4$ in Eq.~\eqref{eq:isoS_expansion} is not equal to $qa$ when $\nu\neq 1$. The condition to saturate the bound~\eqref{eq:multiband_bound} is given by
\begin{align}
    (1-P)(\nabla^2 P)P=0. \label{eq:multiband_saturation}
\end{align}
This is the generalization of the harmonic condition~\eqref{eq:saturate_condition} to multiband cases. When $\nu=1$, Eq.~\eqref{eq:multiband_bound}, \eqref{eq:multiband_saturation} reduce to Eq.~\eqref{eq:geometric_bound}, \eqref{eq:saturate_condition} respectively. 

Importantly, the integrand in Eq.~\eqref{eq:multiband_bound}, which involves a matrix product of two $\mathcal{Q}$, cannot be reduced to the Abelian quantum geometric tensor $\Tr \mathcal{Q}_{\alpha \beta}$. Rather, it is related to a Finsler metric, which measures distance between the points of the Grassmannian manifold associated with the occupied states $\{u_{\vec{k}1},\dots, u_{\vec{k}\nu}\}$~\cite{avdoshkin2024geometry}. 

A simple example of multiband cases saturating the geometric bound is given by an extension of the above one-dimensional tight-binding model to two dimensions. Consider a two-dimensional square lattice with lattice constant $a$ and each site having four orbitals, $\alpha=1,2,3,4$. The hopping is finite only between orbital $1$ at site $\vec{R}$ and $3$ at site $\vec{R}+\vec{a}_{y}$, or between $2$ at site $\vec{R}$ and $4$ at site $\vec{R}+\vec{a}_{x}$, with $\vec{a}_{x(y)}$ a vector with length $a$ in the $x(y)$ direction. The Hamiltonian is given by 
$H = \sum_{\vec{R}} (t_y \cre{c}_1(\vec{R})\ani{c}_3(\vec{R}+\vec{a}_y) + t_x\cre{c}_2(\vec{R})\ani{c}_4(\vec{R}+\vec{a}_x) + \mathrm{h.c.})$,
where $\ani{c}_\alpha(\vec{R})$ is the annihilation operator for orbital $\alpha$ at site $\vec{R}=(na, ma)$.  In this case, the ground state with two electrons per site ($\nu=2$) is given by the collection of the independent bonding orbitals each formed by two orbitals on adjacent sites.

By calculating the quantum geometric tensor directly for this simple model, we obtain the bound on $S_4$ given by the formula Eq.~\eqref{eq:multiband_bound}, 
    $S_4 \ge 3\pi^2/4$.
On the other hand, the static structure factor of this system in real space is given by 
    $S(\vec{R}_1,\vec{R}_2)=\expval{n(\vec{R}_1)n(\vec{R}_2)}-\expval{n(\vec{R}_1)}\expval{n(\vec{R}_2)} = (1/4)\sum_{\vec{d}}(\delta_{\vec{R}_1, \vec{R}_2}-\delta_{\vec{R}_1, \vec{R}_2+\vec{d}})$
where $\vec{d}=\pm\vec{a}_x, \pm\vec{a}_y$ are the vectors connecting the adjacent sites. Then the Fourier transform of $S(\vec{R}_1,\vec{R}_2)$ 
gives the static structure factor $S(\vec{q})=(2-\cos q_x a - \cos q_y a)/4$ which yields $S_4 = 3\pi^2/4$, saturating the geometric bound. %

\textit{$S_4$ bound and quantum geometry fluctuations ---}
The geometric bound on $S_4$ further implies a lower bound in terms of quantum weight and Chern number. To express this relation, we introduce the average of Abelian quantum geometric tensor in $k$-space as $\bar{Q}_{\alpha\beta}\equiv 2\pi\nu^{-1+2/d}\int[\dd{k}]Q_{\alpha\beta}$. Its real part is the quantum weight $K=\Re \bar{Q}$, while the imaginary part yields the Chern number in two dimensions as $C=-2\Im \bar{Q}_{xy}$. 
We can show that the lower bound~\eqref{eq:multiband_bound}, which is given by the quadratic form of quantum geometric tensor $Q$, can itself be bounded with the average $\bar{Q}$ using Cauchy-Schwartz inequality (see SM for details)  
\begin{align}
    S_4 &\ge \frac{12}{d(d+2)}\qty((\tr \bar{Q})^2 + \tr (\bar{Q}^2))\label{eq:geometric_bound2} \\
    &\ge \frac{3(d+1)}{d+2}S_2^2 + \frac{6}{d(d+2)}C_\alpha^2 \label{eq:S4S2_bound}
\end{align}
where $C_{\alpha}=\epsilon_{\alpha\beta\gamma}C_{\beta\gamma}/2$. $C_{z}$ is the Chern number $C$ in $d=2$ and $C_{\alpha}=0$ in $d=1$. The sum over $\alpha=1,\dots, d$ is implied.  
The second inequality follows from relations $\tr(\bar{Q}^2) = \tr (K^2) + C^2_\alpha/2$, $\tr(K^2) \ge (\tr K)^2/d$ and $S_2=2\tr K/d$. Eq.~\eqref{eq:geometric_bound2} is the second main result of this work. It relates $q^4$ term of the structure factor to the leading order in the structure factor and the Chern number. 
It is a direct generalization of Eq.~\eqref{eq:S4_bound_atom}, and applies to general band insulators with arbitrary $\nu$. Notably, the lower bound includes a contribution from the Chern number, which was absent in the atomic insulator example but is nonzero in Chern insulators.  

For two-dimensional Chern insulators, $\tr K\ge \abs{C}$ holds~\cite{roy_band_2014, peotta_superfluidity_2015}, and thus $S_2\ge \abs{C}$ follows. Hence, we find a topological bound on $S_4$:
\begin{align}
    S_4 & \ge 3 C^2. \label{eq:topo_bound}
\end{align}
Therefore, $S_4$ has a universal lower bound determined by topology as $S_2$ does in two dimensions.

For $\nu=1$, the bound~\eqref{eq:geometric_bound2} is saturated only when the band is harmonic \textit{and} the quantum geometric tensor $Q_{\alpha\beta}$ is constant in $k$-space. %
Therefore, the saturation of the topological bound~\eqref{eq:topo_bound} also requires uniform band geometry. The deviation from these bounds thus quantifies fluctuations of the quantum geometry in $k$-space.

\textit{Landau level example. ---}
Notably, both the geometric bound~\eqref{eq:geometric_bound} and topological bound~\eqref{eq:topo_bound} are saturated for the lowest Landau level. For the Landau level with filling factor $\nu=1$, the structure factor is isotropic, given by
\begin{align}
    S(\vec{q}) &= 1-e^{-q^2 l^2/2} 
    = \frac{q^2l^2}{2} - \frac{q^4l^4}{8} + \dots
\end{align}
where $l=(\hbar/eB)^{1/2}$ is the magnetic length. With the carrier density $n=1/(2\pi l^2)$, we find 
    $S_4 = 3$.
Using the quantum metric $g_{\alpha\beta} = \delta_{\alpha\beta}/(4\pi)$ and Berry curvature $\Omega_{xy}=1/(2\pi)$, it is easy to see the lowest Landau level saturates both the geometric and topological bounds.

For the filling factor $\nu=2$, 
the Landau level saturates the geometric bound~\eqref{eq:multiband_bound} but not the topological bound~\eqref{eq:topo_bound} (see SM for details). 
In general, to saturate the topological bound~\eqref{eq:topo_bound} for multiple bands occupied, all the non-Abelian quantum geometric tensor $\mathcal{Q}_{\alpha\beta}$ needs to be proportional to identity. 

Note that the static structure factor of Landau levels we discuss here should be distinguished from the guiding-center structure factor (or the projected structure factor), defined as the correlation function of the projected density operator onto the lowest Landau level. It is known that %
the guiding-center structure factor has a lower bound (Haldane bound) in {\it fractional} quantum Hall states~\cite{haldane_hall_2009}. However, by definition it vanishes identically at {\it integer} fillings. In contrast, our bound is for the \textit{full} static structure factor, which contains a wealth of information about band geometry as we have shown.

So far, we have considered the structure factor for band insulators associated with the entire set of occupied bands. More generally, we can consider the band-resolved structure factor, associated only with a single band $\ket{u_{n\vec{k}}}$, by replacing the projector onto the occupied bands $P$ with $P_n=\ketbra{u_{n\vec{k}}}{u_{n\vec{k}}}$. This band-resolved structure factor also has a geometric bound~\eqref{eq:geometric_bound} determined by the band geometry of $\ket{u_n(\vec{k})}$, and the saturation of the geometric bound defines harmonic bands. 
For example, the band-resolved $S_2$ and $S_4$ for $n$-th Landau level saturate the bound~\eqref{eq:S4S2_bound} for any $n$ (see SM for details), thus they are all harmonic. This contrasts with the trace condition~\cite{roy_band_2014, claassen_position-momentum_2015, ledwith_fractional_2020, mera_kahler_2021, wang_exact_2021, ledwith_family_2022, ledwith_vortexability_2023, liu_theory_2024}, which is satisfied only by the lowest Landau level.

In general, bands satisfying the trace condition are always harmonic, but the converse is not true. To see this, we rewrite the trace condition $\tr g=\abs{\Omega_{xy}}$ as (see Ref.~\cite{claassen_position-momentum_2015, ledwith_fractional_2020, mera_kahler_2021} and SM for details)
\begin{align}
    (D_x\pm iD_y)\ket{u} = 0. \label{eq:trace_condition}
\end{align}
When Eq.~\eqref{eq:trace_condition} is satisfied, Eq.~\eqref{eq:saturate_condition} also holds as seen by multiplying it with $(1-P)(D_x \mp iD_y)$, i.e., the band is harmonic. However, not all harmonic bands satisfy the trace condition. For example, $n\geq 1$ Landau levels are harmonic even though they do not satisfy the trace condition; even a $C=0$ band can be harmonic as shown earlier. 
The harmonic band is thus a broader class than the Chern band satisfying the trace condition. 

Since harmonic Chern bands can be viewed as generalizations of (lowest and higher) Landau levels, the closeness of $S_4$ to the geometric bound~\eqref{eq:S4S2_bound} naturally provides a measure of the similarity between the two. This can inform our search for non-Abelian topological order such as Moore-Read states~\cite{moore_nonabelions_1991, read_paired_2000, levin_particle-hole_2007} and Read-Rezayi state~\cite{read_beyond_1999}, as the prototypical system that hosts non-Abelian topological order is the first-excited Landau level~\cite{willett_observation_1987}.

We also note that the topological bound~\eqref{eq:topo_bound} provides a more stringent condition on Chern bands than the harmonic and trace conditions.
As we discussed earlier, it further requires the band geometry to be uniform in $k$-space.
To illustrate this point, we consider two-dimensional Dirac fermions in a periodic magnetic field $B(\vec{r})=B_0 + B_1(\vec{r})$, where $B_0$ is constant and $B_1(\vec{r})$ is the nonuniform component with zero spatial average.%
 The corresponding Hamiltonian is $H=v_F\vec{\sigma}\vdot(\vec{p}-e\vec{A}(\vec{r}))$ with $\vec{A}(\vec{r})$ the vector potential associated with $B(\vec{r})$. 
This system always hosts a flat Chern band at zero energy that satisfies the trace condition and thus also saturates the topological bound on $S_2$, regardless of spatial magnetic field variation%
~\cite{dong_dirac_2022}. However, strong spatial modulation of the magnetic field results in a highly inhomogeneous local density of states, which should lead to Wigner crystal rather than fractional Chern insulator under realistic Coulomb interaction.  %
This example suggests the limitations of the trace condition in distinguishing ``ideal'' Chern bands and the $n=0$ Landau level. 

\begin{figure}
    \centering
    \includegraphics[width=1.0\columnwidth]{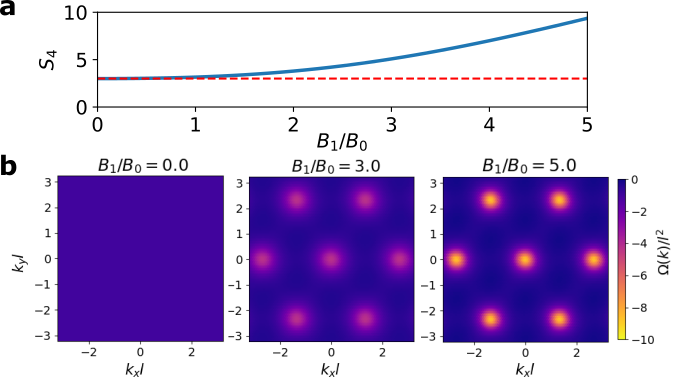}
    \caption{$S_4$ of Dirac fermion model in a periodic magnetic field $B(\vec{r})=B_0 + B_1\sum_{i=1}^3 \cos(\vec{G}_i\vdot\vec{r})$, where $\vec{G}_1 =G(1,0), \vec{G}_2 =G(-1/2, \sqrt{3}/2)$ and $\vec{G}_3 = -\vec{G}_1-\vec{G}_2$, with $G = (1/l)(4\pi/\sqrt{3})^{1/2}$ and $l=\sqrt{\hbar/(eB_0)}$ is the magnetic length associated with the uniform component of the magnetic field $B_0$. (a) $S_4$ as a function of the spatial variation strength $B_1$. The red line is the topological bound~\eqref{eq:topo_bound} with $C=-1$. (b) The Berry curvature distribution in $k$-space. }
    \label{fig:dirac_mag}
\end{figure}

We calculated $S_4$ for this Chern band from a periodic magnetic field. $S_4$ increases with the spatial variation of the magnetic field (Fig.~\ref{fig:dirac_mag}(a)), because the band geometry fluctuates in $k$-space as the magnetic field is spatially modulated (Fig.~\ref{fig:dirac_mag}(b)). 
Therefore, $S_4$ can distinguish different ``ideal'' bands %
even though the trace condition and $S_2$ cannot,  thus offering a more informative guide to identify Chern bands suitable for fractional Chern insulators.

\textit{Conclusion. ---} We have studied the $q^4$ term in the small $q$ expansion of the structure factor $S(\vec{q})$ and showed that it obeys a geometric bound determined by the quadratic form of the quantum geometric tensor in $k$-space. This geometric bound further implies the relation between the fourth-order and second-order terms of $S(\vec{q})$, leading to a topological bound. The deviation from the topological bound quantifies band geometry fluctuations. %

Our results highlight the significance of band geometry fluctuations in $k$-space. %
We showed that the $q^4$ term in the structure factor, directly measurable by scattering experiments, captures the $k$-space distribution of the quantum geometric tensor through its quadratic form. It would be interesting to see if higher-order terms in $S(q)$ can further capture more detailed information on $k$-space distribution of band geometry, and how it can affect other physical quantities.

Furthermore, our new bounds naturally guide us to the idealized systems that saturate the bounds. We see studying such idealized systems as a useful reference for understanding realistic models. This approach can be made more rigorous if one could quantify the deviation from the harmonic limit in a geometrically natural way. We plan to develop this idea in future work.

We thank Timothy Zaklama for participation at the early stage of this project and Daniele Guerci for discussions. 
This work was supported by %
the Air Force Office of Scientific Research under award number FA2386-24-1-4043.
YO is grateful for the support provided by the Funai Overseas Scholarship. 
LF was supported in part by a Simons Investigator Award from the Simons Foundation.

\textit{Note added. --- } Recently, after we posted our manuscript on arXiv, there appeared another work discussing harmonic band and generalized Landau levels~\cite{paiva_geometrical_2025}.

\bibliography{references, ref}

\newpage

\appendix
\begin{widetext}
\newpage
\section*{Supplemental Material}
\section{Derivation of geometric bound in general dimensions}
Here we provide the derivation of the geometric bound in general dimensions $d$. We expand the structure factor $S(\vec{q})$ and its isotropic component $\bar{S}(q)$ in $q$ as follows: 
\begin{align}
    S(\vec{q}) &= \frac{1}{N_e} \expval{n_{\vec{q}}n_{-\vec{q}}} = K_{\alpha\beta}\frac{\bar{q}_{\alpha}\bar{q}_{\beta}}{2\pi}-\frac{L_{\alpha\beta\gamma\delta}}{4!} \frac{\bar{q}_{\alpha}\bar{q}_{\beta}\bar{q}_{\gamma}\bar{q}_{\delta}}{(2\pi)^2} + \dots, \\
    \bar{S}(q) &= \int_{\abs{\vec{q}}=q}\frac{\dd{\Omega}}{A_d}S(\vec{q})=\frac{S_2}{2!}\frac{\bar{q}^2}{2\pi} - \frac{S_4}{4!}\qty(\frac{\bar{q}^2}{2\pi})^2+\dots, 
\end{align}
where $n_{\vec{q}}=\int\dd{\vec{r}}e^{-i\vec{q}\vdot\vec{r}}n(\vec{r})$ is the charge density operator with wavevector $\vec{q}$, $N_e$ is the total number of electrons. Note that $S(\vec{q}\to \infty)=1$ and $S(0)=N_e$. $A_{d}$ is the surface area of the $(d-1)$-dimensional sphere in $d$-dimensions, given by $A_{d}=2\pi^{d/2}/\Gamma(d/2)$ with the Gamma function $\Gamma(z)=\int_0^{\infty}\dd{t}t^{z-1}e^{-z}$. $\bar{q}=qr_0$ is the dimensionless wavevector with $r_0=n^{-1/d}$ the length scale of inter particle distance. $r_0$ is related to $a=V_{\rm uc}^{1/d}$ in the main text as $r_0=a/\nu^{1/d}$ with $\nu$ the number of occupied bands. 

\subsection{Expression for $S_2$ and $S_4$}
Let us first relate $K$ and $L$ to $S_2$ and $S_4$. To this end, we consider the following integral:
\begin{align}
    T(\lambda) &= \int\dd^d\vec{q} \frac{e^{-q^2/(2\lambda^2)}}{(2\pi \lambda^2)^{d/2}} S(\vec{q}) \nonumber \\
    &= A_d\int_0^{\infty}\dd{q} \frac{e^{-q^2/(2\lambda^2)}}{(2\pi \lambda^2)^{d/2}} q^{d-1}\bar{S}(q) \nonumber \\
    &= T_2 \lambda^2 - T_4 \lambda^4 + \dots.
\end{align}
Note that $\lambda^n$ term of $T(\lambda)$, $T_n$, corresponds to $q^n$ term in the static structure factor. One can calculate $\lambda^2$ and $\lambda^4$ terms of $T(\lambda)$ directly from $S(\vec{q})$ as 
\begin{align}
    T_2 &= \frac{r_0^2 K_{\alpha\beta}}{2\pi} \delta_{\alpha\beta}=\frac{r_0^2 K_{\alpha\alpha}}{2\pi}, \\
    T_4 &= \frac{r_0^4L_{\alpha\beta\gamma\delta}}{4! (2\pi)^2} (\delta_{\alpha\beta}\delta_{\gamma\delta} + \delta_{\alpha\gamma}\delta_{\beta\delta} + \delta_{\alpha\delta}\delta_{\beta\gamma}) \nonumber \\
    &= \frac{r_0^4}{4! (2\pi)^2} (L_{\alpha\alpha\beta\beta} + L_{\alpha\beta\alpha\beta} + L_{\alpha\beta\beta\alpha})
\end{align}
On the other hand, $T_2, T_4$ can be also calculated from $\bar{S}(q)$ as 
\begin{align}
    T_2 &= \frac{S_2 r_0^2}{2! (2\pi)} 2 \frac{\Gamma(1+d/2)}{\Gamma(d/2)}= \frac{S_2 r_0^2}{2! (2\pi)}d\\
    T_4 &= \frac{S_4 r_0^4}{4! (2\pi)^2} 2^2 \frac{\Gamma(2+d/2)}{\Gamma(d/2)} = \frac{S_4 r_0^4}{4! (2\pi)^2} d(d+2) 
\end{align}
Comparing these expressions, we find the relation between $K, L$ and $S_2, S_4$ as 
\begin{align}
    S_2 &= \frac{2}{d}K_{\alpha\alpha} \label{ap:eq:S2_K} \\
    S_4 &= \frac{L_{\alpha\alpha\beta\beta} + L_{\alpha\beta\alpha\beta} + L_{\alpha\beta\beta\alpha}}{d(d+2)} \label{ap:eq:S4_L}
\end{align}

Now let us calculate $S_2$ and $S_4$ for band insulators. In this case, the static structure factor is given by 
\begin{align}
    S(\vec{q}) 
    &= \frac{1}{\nu}\int[dk]\Tr[P(\vec{k})(P(\vec{k})-P(\vec{k}+\vec{q}))] 
    \label{ap:eq:Sq}
\end{align}
where $N_e$ is the total number of electrons, and $P(\vec{k})=\sum_{n=1}^{\nu}\ketbra{u_n(\vec{k})}{u_n(\vec{k})}$ is the projector onto the occupied bands at wavevector $\vec{k}$ with the {\it cell-periodic} Bloch wavefunction of $n$-th band $\ket{u_n(\vec{k})}$ and the number of occupied bands $\nu$. $\int[dk] = \int_{\rm BZ} V_{\rm uc}d^dk/(2\pi)^d$ is the integral over the Brillouin zone normalized with the unit cell area $V_{\rm uc}$. We can confirm $S(\vec{q})\to 1$ at $\vec{q}\to \infty$ since $\Tr[P(\vec{k})P(\vec{k}+\vec{q})]\to 0$ in this limit and $\Tr[P(\vec{k})^2]=\Tr[P(\vec{k})]=\nu$. 

By expanding Eq.~\eqref{ap:eq:Sq}, we find that $K$ and $L$ as 
\begin{align}
    K_{\alpha\beta} &= \frac{\pi}{\nu r_0^2}\int[dk]\Tr[(\partial_{\alpha}P(\vec{k}))(\partial_{\beta}P(\vec{k}))] \\
    L_{\alpha\beta\gamma\delta} &= \frac{(2\pi)^2}{\nu r_0^4}\int[dk]\Tr[(\partial_{\alpha}\partial_{\beta} P)(\partial_{\gamma}\partial_{\delta} P)]  
\end{align}
From these expressions and Eq.~\eqref{ap:eq:S2_K} and \eqref{ap:eq:S4_L}, we obtain $S_2$ and $S_4$ as 
\begin{align}
    S_2 &= \frac{2\pi}{d \nu r_0^2}\int[dk]\Tr[(\partial_{\alpha} P)(\partial_{\alpha} P)]\\
    S_4 &= \frac{3(2\pi)^2}{d(d+2)\nu r_0^4}\int[dk]\Tr[(\nabla^2 P)^2]
\end{align}

\subsection{Bound on $S_4$}
\subsubsection{For $\nu=1$}
For $\nu=1$, $S_4$ is given by
\begin{align}
    S_4 &= \frac{3(2\pi)^2}{d(d+2)a^4}\int[\dd{k}] \Tr[(\nabla^2 P)^2]. \label{ap:q:S4_2dim}
\end{align}
Using the Bloch wavefunction, we can rewrite $\nabla^2 P$ as:
\begin{align}
    \nabla^2 P &= \ketbra{D^2u}{u} + 2\ketbra{D_{\alpha}u}{D_{\alpha}u} + \ketbra{u}{D^2 u} 
\end{align}
With this expression and relations $\ip{u}{D^2u}=a^2\tr Q$ and $\ip{u}{D_{\alpha}u} = 0$, $\Tr[(\nabla^2P)^2]$ can be rewritten as 
\begin{align}
    \frac{1}{a^4}\Tr[(\nabla^2 P)^2] 
    &= 4\qty((\tr Q)^2 + \tr (Q^2)) + \frac{2}{a^4} \mel{D^2 u}{(1-P)}{D^2 u}.  
\end{align}
With this expression, we can rewrite $S_4$ as follows:
\begin{align}
    S_4 
    &= \frac{12(2\pi)^2}{d(d+2)}\int[\dd{k}] \qty((\tr Q)^2 + \tr (Q^2)) + \frac{6(2\pi)^2}{d(d+2)a^4}\int[\dd{k}] \mel{D^2 u}{(1-P)}{D^2 u}. \label{ap:eq:S4_general}
\end{align}
This is the expression of $S_4$ for $\nu=1$ in the main text. As discussed in the main text, the last term is always semi-positive, thus we obtained the geometric bound on $S_4$ as 
\begin{align}
    S_4 &\ge \frac{12(2\pi)^2}{d(d+2)}\int[\dd{k}] \qty((\tr Q)^2 + \tr (Q^2))
\end{align}

\subsubsection{For general $\nu$}
Now we show the bound on $S_4$ in terms of quantum geometry. First, we define the quantum geometric tensor as follows:
\begin{align}
    \mathcal{Q}_{\alpha\beta}\equiv \frac{1}{a^2}(\partial_{\alpha} P)(1-P)(\partial_{\beta}P)
\end{align}
Note that $(\mathcal{Q}_{\alpha\beta})^{\dagger}=\mathcal{Q}_{\beta\alpha}$..
From $\mathcal{Q}$, we can also define the non-Abelian quantum metric $\mathcal{G}$ and Berry curvature $\mathcal{F}$ as  $\mathcal{G}_{\alpha\beta}\equiv(\mathcal{Q}_{\alpha\beta}+\mathcal{Q}_{\beta\alpha})/2$ and $\mathcal{F}_{\alpha\beta}\equiv i(\mathcal{Q}_{\alpha\beta}-\mathcal{Q}_{\beta\alpha})$, respectively. $\mathcal{G}_{\alpha\beta}, \mathcal{F}_{\alpha\beta}$ are hermitian. 

With the quantum geometric tensor, we can  rewrite $a^4\Tr[(\nabla^2 P)^2]$ as follows. 
$\Tr[(\nabla^2 P)^2]$ can be written as 
\begin{align}
    \Tr[(\nabla^2 P)^2] &= \Tr[(\nabla^2 P)(P+P^c)(\nabla^2 P)(P+P^c)] \nonumber \\
    &= \Tr[(\nabla^2 P)P(\nabla^2 P)P + (\nabla^2 P)P^c(\nabla^2 P)P^c + 2(\nabla^2 P)P(\nabla^2 P)P^c] \label{ap:eq:nablaP2}
\end{align}
where $P^c = 1-P$. To proceed, the following relations are useful:
\begin{align}
    0 &= (\nabla P)P^c - P \nabla P = P^c(\nabla P)- (\nabla P) P \\
    0 &= (\nabla^2 P) P^c - 2(\nabla P)^2 - P(\nabla P)^2
\end{align}
These relations can be shown by taking the derivative of $PP^c = P^cP = 0$.
With these, we can show the following:
\begin{align}
    P(\nabla^2 P) P 
    &= -2 (\nabla P)P^c(\nabla P) \\
    P^c(\nabla^2 P)P^c &= 
    2P^c (\nabla P)^2 
\end{align}
Therefore, Eq.~\eqref{ap:eq:nablaP2} can be rewritten as 
\begin{align}
    \Tr[(\nabla^2 P)^2] &= \Tr[(P(\nabla^2 P)P)^2 + (P^c(\nabla^2 P)P^c)^2 + 2P(\nabla^2 P)P^c(\nabla^2 P)P] \nonumber \\
    &= \Tr[4 ((\nabla P)P^c (\nabla P))^2 + 4 (P^c(\nabla P)^2)^2] + 2\Tr[P(\nabla^2 P) P^c (\nabla^2 P) P] \nonumber \\
    &= 4a^{4}\Tr[\mathcal{Q}_{\alpha\alpha}\mathcal{Q}_{\beta\beta}+\mathcal{Q}_{\alpha\beta}\mathcal{Q}_{\beta\alpha}] + 2\Tr[P(\nabla^2 P) P^c (\nabla^2 P) P]
\end{align}
with $\mathcal{Q}$ the non-Abelian geometric tensor defined above.
Since the second term is semi-positive, we find the lower bound on $S_4$ as 
\begin{align}
    S_4 &\ge \frac{12(2\pi)^2 \nu^{-1+(4/d)}}{d(d+2)}\sum_{\alpha,\beta}\int[dk]\Tr[\mathcal{Q}_{\alpha\alpha}\mathcal{Q}_{\beta\beta}+\mathcal{Q}_{\alpha\beta}\mathcal{Q}_{\beta\alpha}] \label{ap:eq:S4_QQ} 
\end{align}
This bound is saturated when the following condition is satisfied:
\begin{align}
    (1-P)(\nabla^2 P)P = 0. \label{ap:nonabel_harmonic_condition}
\end{align}
This is the non-Abelian version of the harmonic condition. 

One can further rewrite Eq.~\eqref{ap:eq:S4_QQ} in terms of $q^2$ coefficient of $S(\vec{q})$, $K$, and the Chern number. To this end, it is convenient to introduce the average of the Abelian quantum geometric tensor $\bar{Q}$ as 
\begin{align}
    \bar{Q}_{\alpha\beta}=2\pi\nu^{-1+(2/d)}\int[\dd k] Q_{\alpha\beta} = 2\pi\nu^{-1+(2/d)}\int[\dd k] \Tr[\mathcal{Q}_{\alpha\beta}].
\end{align}
where $Q_{\alpha\beta}\equiv \Tr\mathcal{Q}_{\alpha\beta} = g_{\alpha\beta}-(i/2)\Omega_{\alpha\beta}$ is the abelian quantum geometric tensor and $g, \Omega$ are the quantum metric and the Berry curvature.
Then $K=2\pi\nu^{-1+(2/d)}\int [\dd k] g_{\alpha\beta}$ is given by the real part of $\bar{Q}$. 
Correspondingly, we define a dimensionless rank-2 tensor $C$ as $C=-2\Im \bar{Q}$. For two dimensions, $C_{xy}$ reduces to the Chern number.
With $K$ and $C$, we can find a lower bound on the integral in Eq.~\eqref{ap:eq:S4_QQ} as follows. First, we note the following inequality:
\begin{align}
    \Tr[\mathcal{Q}_{\alpha\alpha}\mathcal{Q}_{\beta\beta}+\mathcal{Q}_{\alpha\beta}\mathcal{Q}_{\beta\alpha}] &\ge \frac{1}{\Tr[P^2]}(\Tr[\mathcal{Q}_{\alpha\alpha} P] \Tr[\mathcal{Q}_{\beta\beta} P] + \Tr[\mathcal{Q}_{\alpha\beta} P]\Tr[\mathcal{Q}_{\beta\alpha}P]) 
\end{align}
where we have used the Cauchy-Schwartz inequality for the matrices, $(\Tr[A^{\dagger}B])^2 \le \Tr[A^\dagger A]\Tr[B^\dagger B]$ with $A=\mathcal{Q}_{\alpha\alpha}, \mathcal{Q}_{\alpha\beta}$ and $B=P$. Noting that $\Tr[P^2]=\Tr[P] = \nu$ and $\mathcal{Q}_{\alpha\beta}P=\mathcal{Q}_{\alpha\beta}$, we find 
\begin{align}
    S_4 &\ge \frac{12(2\pi \nu^{-1+(2/d)})^2}{d(d+2)}\int[\dd{k}] \qty(Q_{\alpha\alpha}Q_{\beta\beta}+Q_{\alpha\beta}Q_{\beta\alpha}).
\end{align}
This inequality is saturated only when the non-Abelian harmonic condition~\eqref{ap:nonabel_harmonic_condition} is satisfied and $\mathcal{Q}_{\alpha\beta}$ is proportional to $P$ for all $\alpha$ and $\beta$. 

Further using that Cauchy-Schwartz inequality for the integral, $\abs{\int f^*(\vec{k}) g(\vec{k}) [\dd{k}]}^2 \le \int \abs{f(\vec{k})}^2 [\dd{k}] \int\abs{g(\vec{k})}^2 [\dd{k}]$ with $f(\vec{k})=Q_{\alpha\alpha}$ or $Q_{\alpha\beta}$ and $g(\vec{k})=1$, we find 
\begin{align}
    S_4 &\ge 
    \frac{12}{d(d+2)}\qty((\tr\bar{Q})^2 + \tr(\bar{Q}^2)) \\
    &= \frac{12}{d(d+2)}\qty((\tr K)^2 + \tr((K- iC/2)^2))
\end{align}
where $\bar{Q}=2\pi\nu^{-1+(2/d)}\int[\dd k] Q$. This inequality is saturated when the non-Abelian harmonic condition~\eqref{ap:nonabel_harmonic_condition} is satisfied, $\mathcal{Q}_{\alpha\beta}$ is proportional to $P$ for all $\alpha$ and $\beta$, and the Abeliang geometric tensor $Q_{\alpha\beta}$ is constant in $k$-space. 

Since $\tr(K^2) \ge (\tr K)^2/d$ and $S_2 = \tr K$, we have 
\begin{align}
    S_4 
    &\ge \frac{3 (d+1)}{d+2}(S_2)^2 + \frac{6 C_{\alpha}^2}{d(d+2)}
\end{align}
with $C_{\alpha}=\epsilon_{\alpha\beta\gamma}C_{\beta\gamma}/2$.

In particular, for $d=1,2,3$, 
\begin{align}
     S_4 &\ge 2(S_2)^2 &&\text{($d=1$)} \\
     S_4 &\ge \frac{9}{4}(S_2)^2 + \frac{3}{4}C^2 &&\text{($d=2$)} \\
     S_4 &\ge \frac{12}{5}(S_2)^2 + \frac{2}{5}(C_{x}^2 + C_y^2 + C_z^2) &&\text{($d=3$)}
\end{align}

\section{Atomic insulators}
Here we provide a detailed calculation of the quantum geometry and structure factor of atomic insulators.

Consider a periodic array of atoms far away from each other with lattice constant $a$. When $a$ is large enough, the energy eigenstate is identical to the one for each atom, $\ket{\phi_n(\vec{r}-\vec{R})}$, where $\vec{R}$ is the position of an atom and $n$ is the label for each atomic orbital. Then the Bloch Hamiltonian is given by 
\begin{align}
    \psi_{n\vec{k}}(\vec{r}) &= \frac{1}{\sqrt{N}}\sum_{\vec{R}} e^{i\vec{k}\vdot(\vec{R}+\vec{r}_n)}\phi_n(\vec{r}-\vec{R}) \\
    u_{n\vec{k}}(\vec{r}) &= e^{-i\vec{k}\vdot\vec{r}}\psi_{n\vec{k}}(\vec{r}) = \frac{1}{\sqrt{N}}\sum_{\vec{R}} e^{-i\vec{k}\vdot(\vec{r}-\vec{R}-\vec{r}_n)}\phi_n(\vec{r}-\vec{R})
\end{align}
where $\vec{r}_n = \int\abs{\phi_n(\vec{r})}^2\dd{\vec{r}}$ is the expectation value of the position for each atomic orbital $\phi_n(\vec{r})$. Note that $u_{n\vec{k}}(\vec{r})$ is $k$-dependent even in the atomic limit.

\subsection{Quantum geometric tensor of atomic insulators}
Quantum geometric tensor for $n$-th band is defined as 
\begin{align}
    Q_{\alpha\beta}(\vec{k}) &= \frac{1}{a^2}\mel{\partial_{\alpha}u_{n\vec{k}}}{(1-P)}{\partial_{\beta}u_{n\vec{k}}}
\end{align}
For the atomic limit, the quantum geometric tensor is given by 
\begin{align}
    Q_{\alpha\beta}(\vec{k}) 
    &= \frac{1}{a^2}\int \dd{r} \phi_n^*(\vec{r}) (\vec{r}-\vec{r}_n)_{\alpha}(\vec{r}-\vec{r}_n)_{\beta}\phi_n(\vec{r}) \\
    &= \frac{1}{a^2}\expval{\Delta r_{\alpha}\Delta r_{\beta}}_n
\end{align}
where $\Delta\vec{r}=\vec{r}-\vec{r}_n$. 
Therefore the quantum geometric tensor is always real and corresponds to the position fluctuation.
The quantum metric $g_{\alpha\beta}=\Re Q_{\alpha\beta}$ and the Berry curvature $\Omega_{\alpha\beta}=-2\Im Q_{\alpha\beta}$ is thus 
\begin{align}
    g_{\alpha\beta} &= \expval{\Delta r_{\alpha}\Delta r_{\beta}}_n \\
    \Omega_{\alpha\beta} &= 0.
\end{align}

\subsection{Structure factor of atomic insulators}
To calculate the structure factor in this system, it is convenient to use the expression for the structure factor in real space (for the derivation, see Supplemental Material of Ref.~\cite{onishi2024h}). For noninteracting systems, the equal-time density-density correlation function $S(\vec{r}_1, \vec{r}_2)\equiv \expval{n(\vec{r}_1)n(\vec{r}_2)}-\expval{n(\vec{r}_1)}\expval{n(\vec{r}_2)}$ is given by 
\begin{align}
    S(\vec{r}_1, \vec{r}_2) &= (\delta(\vec{r}_1,\vec{r}_2)-P(\vec{r}_1, \vec{r}_2)) P(\vec{r}_2, \vec{r}_1). \label{ap:eq:Sq_P}
\end{align}
$P(\vec{r}_1, \vec{r}_2)=\mel{\vec{r}_1}{P}{\vec{r}_2}$ is the matrix element of the projection operator onto the occupied states, given by 
\begin{align}
    P &= \sum_{a:{\rm occ}}\ketbra{\psi_a}{\psi_a},  \\
    P(\vec{r}_1, \vec{r}_2) &= \sum_{a:{\rm occ}}\psi_a(\vec{r}_1)\psi_a^*(\vec{r}_2)
\end{align}
where $\psi_a$ is a normalized occupied state orthogonal to each other, i.e., $\ip{\psi_a}{\psi_b}=\int_V \psi_a^*(\vec{r})\psi_b(\vec{r}) =\delta_{ab}$ with integration taken to be over the entire system volume $V$. Accordingly, $\int_V P(\vec{r}, \vec{r})\dd{r} = N_e$ with $N_e$ the total number of electrons. 

From the density-density correlation function in real space $S(\vec{r}_1, \vec{r}_2)$, we can obtain its Fourier transform $S'_{\vec{q}}$ as 
\begin{align}
    S'_{\vec{q}} &= \frac{1}{N_e}\int\dd{\vec{r}_1}\dd{\vec{r}_2} e^{-i\vec{q}\vdot(\vec{r}_1-\vec{r}_2)} S(\vec{r}_1, \vec{r}_2) \label{eq:Sq_Sr}
\end{align}
where $S'_{\vec{q}}=(1/N_e)(\expval{n_{\vec{q}}n_{-\vec{q}}}-\expval{n_{\vec{q}}}\expval{n_{-\vec{q}}})$ and $N_e$ is the total number of electrons. Note that $S'_{\vec{q}}$ is identical with $S(\vec{q})$ at $\vec{q}$ away from the reciprocal lattice vectors, since $\expval{n_{\vec{q}}}=0$ for such $\vec{q}$.

With these expressions, we can easily calculate the structure factor for the atomic insulator. When each atom has one electron occupying $n=0$ orbital, the projection operator $P$ is given by $P=\sum_{\vec{R}}\ketbra{\phi_0(\vec{r}-\vec{R})}{\phi_0(\vec{r}-\vec{R})}$, where $\vec{R}$ specifies a unit cell. Then the structure factor $S'_{\vec{q}}$ is 
\begin{align}
    S'_{\vec{q}} &= \frac{1}{N_e}\int\dd{r} \sum_{\vec{R}}\abs{\phi_0(\vec{r}-\vec{R})}^2 - \frac{1}{N_e}\sum_{\vec{R}_1, \vec{R}_2}\int\dd{\vec{r}_1}e^{-i\vec{q}\vdot\vec{r}_1} \phi_0(\vec{r}_1-\vec{R}_1)\phi_0^*(\vec{r}_1-\vec{R}_2)\int\dd{\vec{r}_2}e^{i\vec{q}\vdot\vec{r}_2} \phi_0(\vec{r}_2-\vec{R}_1)\phi_0^*(\vec{r}_2-\vec{R}_2) \nonumber \\
    &= 1 - \frac{1}{N_e}\sum_{\vec{R}}\abs{\int\dd{\vec{r}}e^{-i\vec{q}\vdot\vec{r}} \abs{\phi_0(\vec{r}-\vec{R})}^2}^2 = 1 - \abs{\int\dd{\vec{r}}e^{-i\vec{q}\vdot\vec{r}} \abs{\phi_0(\vec{r})}^2}^2 = 1-\abs{\expval{e^{-i\vec{q}\vdot\vec{r}}}}^2 \label{ap:eq:structure_factor}
\end{align}
where $\expval{\dots}=\int\dd{r}(\dots)\abs{\phi_0(\vec{r})}^2$ is the expectation value in state $\phi_0(\vec{r})$. Here, we assumed $\phi_0(\vec{r}-\vec{R}_1)\phi_0^*(\vec{r}-\vec{R}_2)$ is finite only when $\vec{R}_1=\vec{R}_2$ because $\phi_0$ is localized compared to the lattice constant $a$. Since $S'_{\vec{q}}$ is identical with $S(\vec{q})$ at $\vec{q}$ away from the reciprocal lattice vectors, Eq.~\eqref{ap:eq:structure_factor} gives the expression for the structure factor in the main text.

By further expanding in $\vec{q}$, we obtain
\begin{align}
    S(\vec{q}) 
    &= \expval{(\vec{q}\vdot\vec{r})^2} - \frac{2}{4!}\qty(\expval{(\vec{q}\vdot\vec{r})^4} + 3\expval{(\vec{q}\vdot\vec{r})^2}^2) + \dots. \nonumber \\
    &= \frac{K_{\alpha\beta}}{2\pi}q_{\alpha}q_{\beta}a^2 - \frac{L_{\alpha\beta\gamma\delta}}{4!} q_{\alpha}q_{\beta}q_{\gamma}q_{\delta}a^4 + \dots,
\end{align}
where the coefficients $K$ and $L^S$ are given by
\begin{align}
    K_{\alpha\beta} &= 2\pi \expval{r_{\alpha}r_{\beta}}/a^2, \\
    L_{\alpha\beta\gamma\delta} &= 2(\expval{r_{\alpha}r_{\beta}r_{\gamma}r_{\delta}} + \expval{r_{\alpha}r_{\beta}}\expval{r_{\gamma}r_{\delta}} \nonumber \\
    &+ \expval{r_{\alpha}r_{\gamma}}\expval{r_{\beta}r_{\delta}} + 
    \expval{r_{\alpha}r_{\delta}}\expval{r_{\beta}r_{\gamma}})/a^4.
\end{align}
Note that $a=r_0=n^{-1/d}$ in this case since $\nu=1$. Here, we have chosen the origin so that $\expval{\vec{r}}=0$.

In particular, the isotropic component of $S(\vec{q})$ defined as $\isoS(\vec{q})=\int\dd{\theta} S(q\cos\theta, q\sin\theta)/(2\pi)$ is given by  
\begin{align}
    \isoS(\vec{q}) &= \frac{S_2}{2!}\frac{q^2a^2}{2\pi} - \frac{S_4}{4!}\qty(\frac{q^2a^2}{2\pi})^2 + \dots, 
\end{align}
where, from Eq.~\eqref{ap:eq:S2_K} and \eqref{ap:eq:S4_L}, $S_2$ and $S_4$ are given by 
\begin{align}
    S_2 &= \frac{4\pi}{da^2}\expval{r^2},\\
    S_4 
    &= \frac{12}{d(d+2)} \qty((\tr K)^2 + \tr (K^2)) + \frac{6 (2\pi)^2}{d(d+2) a^4} \expval{(r^2-\expval{r^2})^2}_0
\end{align}
\section{Trace condition}
The trace condition $\tr g = \abs{\Omega_{xy}}$ is equivalent to that the Bloch wavefunction $\ket{u_{\vec{k}}}$ satisfies one of the following conditions:
\begin{align}
    (D_x \pm i D_y)\ket{u}=0. \label{ap:eq:trace}
\end{align}
Mathematically equivalent expressions are found in Ref.~\cite{claassen_position-momentum_2015, ledwith_fractional_2020, mera_kahler_2021}. Here we provide the proof for completeness. 

First we note that when $\tr g=\pm{\Omega_{xy}}$, an inequality $\tr g\ge 2\sqrt{\det g} \ge \abs{\Omega_{xy}}$ is saturated, which implies that $\tr g = 2\sqrt{\det g}$. This means that $g$ is proportional to identity: $g_{\alpha\beta}=g_{xx} \delta_{\alpha\beta}$.  Therefore, the trace condition is equivalent to a condition $g_{xx}=g_{yy}=\pm{\Omega_{xy}}/2$, and the quantum geometric tensor $Q$ takes the form $Q_{\alpha\beta} = g_{xx}(\delta_{\alpha\beta} \pm i\epsilon_{\alpha\beta})$,  with $\epsilon_{\alpha\beta}$ an antisymmetric tensor.
Then the Bloch wavefunction satisfies the following conditions:
\begin{align}
    \ip{D_x u}{D_x u} = \ip{D_y u}{D_y u} = \pm i\ip{D_x u}{D_y u}.
\end{align}
This condition implies that $\ket{D_x u}$ and $\ket{D_y u}$ are parallel and have the same length, and are related to each other by a factor of $\pm i$. Therefore, it follows that 
\begin{align}
    \ket{D_x u} = \pm i \ket{D_y u}
\end{align}
which yields Eq.~\eqref{ap:eq:trace}. 

We also note that Eq.~\eqref{ap:eq:trace} can be written as $(1-P)(\partial_x \pm i\partial_y)P=0$ with the projector $P=\ketbra{u}{u}$. This expression appears in Ref.~\cite{mera_kahler_2021}. 

As discussed in the main text, when a band is ideal and satisfies Eq.~\eqref{ap:eq:trace}, the band is also harmonic. This can be seen by multiplying $(1-P)(D_x\mp iD_y)$ to Eq.~\eqref{ap:eq:trace} and use $\comm{D_x}{D_y}=i(\partial_x A_y -\partial_y A_x)=\Omega_{xy}$ to obtain 
\begin{align}
    (1-P)(D_x^2 + D_y^2 \mp \Omega_{xy}) \ket{u} = (1-P)(D_x^2 + D_y^2) \ket{u} = 0.
\end{align}
This is nothing but the condition for a band to be harmonic.
\section{Landau level}
\subsection{Structure factor of Landau level}
Here we summarize the calculation of the structure factor of the Landau level known in the literature (see, for example, Chapter 10 of Ref.~\cite{giuliani_quantum_2005}).

For the structure factor calculation of the Landau level, the formula in terms of the real space~\eqref{ap:eq:Sq_P} is useful. 

In the Landau gauge, the wavefunction for the $n$-th Landau level ($n=0$ is the lowest Landau level) is given by 
\begin{align}
    \psi_{n,k_y}(\vec{r}) &= \frac{1}{\sqrt{L_y}}e^{ik_y y}\phi_{n, k_y}(x), \\
    \phi_{n, k_y}(x) &= \frac{1}{\sqrt{2^n n! \pi^{1/2}}l} e^{-(x-k_yl^2)^2/(2l^2)} H_n\qty(\frac{x-k_yl^2}{l}), \label{ap:eq:LL_wf}
\end{align}
where $l=\hbar/(eB)$ is the magnetic length, $L_y$ is the system size in $y$-direction, and $H_n(x)$ is the Hermite polynomial of order $n$. $eB>0$ is assumed for simplicity. 

From this, we can calculate the projection operator for the $n$-th Landau level $P_n(\vec{r}, \vec{r}')$. (In Ref.~\cite{giuliani_quantum_2005}, the projection operator here is called a one-particle density matrix.) $P_n(\vec{r}, \vec{r}')$ is given by 
\begin{align}
    P_n(\vec{r}, \vec{r}') &= \sum_{k_y} \psi_{n,k_y}^*(\vec{r})\psi_{n,k_y}(\vec{r}') = \frac{1}{2\pi l^2} e^{-i\frac{(x+x')(y-y')}{2l^2}}e^{-\frac{\abs{\vec{r}-\vec{r}'}^2}{4l^2}} L^0_{n}\qty(\frac{\abs{\vec{r}-\vec{r}'}^2}{2l^2}),
\end{align}
where $L^\alpha_n(x)$ is the Laguerre polynomial defined as 
\begin{align}
    L^{\alpha}_n(x) &= \frac{1}{n!}e^x x^{-\alpha}\dv[n]{x}(e^{-x}x^{n+\alpha})
\end{align}
Note that $P_n(\vec{r}, \vec{r}) = 1/(2\pi l^2)$.
The projection operator~\eqref{ap:eq:LL_wf} coincides with the projection operator for symmetric gauge by gauge transformation $P(\vec{r}, \vec{r}')\to e^{i\chi(\vec{r})}P(\vec{r}, \vec{r}')e^{-i\chi(\vec{r}')}$ with $\chi(\vec{r})=Bxy/2$.

Then with the projection operator for the occupied bands $P=\sum_{n}^{\rm occ}P_n(\vec{r}, \vec{r}')$, the structure factor is given by the Fourier transform of Eq.~\eqref{ap:eq:Sq_P}. 

When the filling factor $\nu=1$, only the lowest Landau level ($n=0$) is filled and $P=P_0$. Then the structure factor is given by
\begin{align}
    S(\vec{q}) &= 1-e^{-q^2 l^2/2} = \frac{q^2 l^2}{2} - \frac{q^4 l^4}{8} + \dots, \quad (\text{$\nu=1$})
\end{align}
Noting that $A=2\pi l^2$, we have $S_2 = 1$, $S_4 = 3$. 

We can also think of the band-resolved structure factor for the second lowest Landau level ($n=1$). Then $P=P_1$ and the band-resolved structure factor is 
\begin{align}
    S(\vec{q}) &= 1-e^{-q^2 l^2/2} \qty(1-\frac{q^2 l^2}{2})^2 = \frac{3q^2 l^2}{2} - \frac{7q^4 l^4}{8} + \dots, \quad (\text{band-resolved structure for $n=1$})
\end{align}
Therefore, $S_2 = 3, S_4 = 21$. More generally, the band-resolved structure factor for $n$-th Landau level is given by
\begin{align}
    S_n^{\rm LL}(\vec{q}) &= 1-\frac{1}{\pi}\int\dd^2{\vec{x}}  e^{-\sqrt{2}li\vec{q}\vdot\vec{x}}e^{-x^2} \abs{L^0_{n}\qty(x^2)}^2
\end{align}
By expanding this expression, we can calculate $q^2$-term and $q^4$-term. $S_2$ and $S_4$ are then given by 
\begin{align}
    S_2 = 2n+1, \\
    S_4 = 3(3n^2 + 3n + 1).
\end{align}
With $C=1$, one can confirm that the geometric bound in two dimensions $S_4\ge 9S_2^2/4 + (3/4)$ is saturated for the band-resolved structure factor for general $n$-th Landau levels. Therefore, each Landau level is a harmonic band. 

When the filling factor $\nu=2$ so that $n=0$ and $n=1$ are filled, the structure factor is given by 
\begin{align}
    S(\vec{q}) &= 1-e^{-q^2l^2/2}\qty(1+\frac{q^2l^2}{8}) = \frac{q^2 l^2}{2}-\frac{q^4l^2}{4} + \dots
\end{align}
Noting that, since $n=\nu/(2\pi l^2)$ and thus $r_0=n^{-1/d}=\sqrt{2\pi l^2/\nu}$, $\bar{q}^2/(2\pi)=q^2r_0^2/(2\pi)=q^2l^2/\nu$. Therefore, $S_2$ and $S_4$ are given by $S_2=2$, $S_4=24$.

\subsection{Quantum geometry of Landau level}
Here we summarize the calculation of the quantum geometry of the Landau level. The Abelian quantum geometric tensor of Landau level for general filling factor $\nu$ was calculated in Ref.~\cite{ozawa_relations_2021}. Here, we show that the $n=0$ and $n=1$ Landau levels are harmonic bands, and also provide the non-Abelian quantum geometric tensor for $\nu=2$ by extending the calculation in Ref.~\cite{ozawa_relations_2021}. 

To calculate the quantum geometry, we first need to construct the Bloch wavefunction that respects the magnetic translational symmetry defined as follows. We first choose a unit cell of a rectangle $a_x \times a_y$, satisfying $a_x a_y = 2\pi l^2$ with $l=\sqrt{\hbar/(eB)}$ the magnetic length. Note that the choice of $a_x, a_y$ does not affect the results below. Then the magnetic translation operators of $\vec{a}_x\equiv a_x \hat{x}, \vec{a}_y \equiv a_y\hat{y}$ are defined as $M(\vec{a}_x)=e^{ia_x y/l^2} T(\vec{a}_x), M(\vec{a}_y)=T(\vec{a}_y)$ with the translation operator by $\vec{R}$ denoted as $T(\vec{R})$. When $a_xa_y = 2\pi l^2$, $M(\vec{a}_x)$ and $M(\vec{a}_y)$ commutes with each other, and the Bloch wavefunction for $n$-th Landau level $\ket{\psi_{n\vec{k}}}$ is chosen so that it is an eigenstate of $M(\vec{a}_x), M(\vec{a}_y)$. Accordingly, the cell-periodic Bloch wavefunction is defined as $\ket{u_{n\vec{k}}}\equiv e^{-i\vec{k}\vdot\vec{r}}\ket{\psi_{n\vec{k}}}$. This procedure was done in Ref.~\cite{ozawa_relations_2021}, and the obtained cell-periodic Bloch wavefunction is given by 
\begin{align}
    u_{n\vec{k}}(\vec{r}) &= \sum_{m=-\infty}^{\infty} e^{ik_x (a_x m -x)} e^{i 2\pi m y/a_y} \varphi_n\qty(x-\frac{\hbar}{eB}(k_y+2\pi m/a_y)), \label{ap:eq:Bloch_LL}\\
    \varphi_n(x) &\equiv \frac{1}{\sqrt{2^n n! \pi^{1/2}}} e^{-x^2/(2l^2)} H_n\qty(x/l), \label{ap:eq:varphi}
\end{align}
where $H_n(x)$ is the Hermite polynomial of degree $n$. Useful formulas of $\varphi_n(x)$ are summarized in Sec.~\ref{ap:sec:formulas}. It can be verified that $u_{n\vec{k}}(\vec{r})$ is normalized as follows:
\begin{align}
    \int_{\rm uc}(u_{n\vec{k}}(\vec{r}))^*u_{m\vec{k}}(\vec{r}) \dd^2{r} = \delta_{nm}
\end{align}
where $\int_{\rm uc}\dd^2{r}$ denotes the integral over one magnetic unit cell. 

With these, we can calculate the quantum geometric quantities of the Landau level. Defining the quantum geometric tensor for $n$-th Landau level as $Q_{n,\alpha\beta}=(1/A)\mel{\partial_{\alpha}u_{n\vec{k}}}{(1-P)}{\partial_{\beta}u_{n\vec{k}}}$ with $A=2\pi l^2$ the unit cell area, the quantum metric $g_{n}=\Re Q_{n}$, and the Berry curvature $\Omega_{n}=-2\Im Q_{n}$ with are given by
\begin{align}
    g_{n,\alpha\beta} &= \frac{1}{2\pi} (n+1/2) \delta_{\alpha\beta} \\
    \Omega_{n, xy} &= \frac{1}{2\pi} 
\end{align}

The non-Abelian quantum geometric tensor can be also calculated with the formulas in Sec.~\ref{ap:sec:formulas}. For $n$-th and $n+1$ Landau levels, the non-Abelian geometric tensor is given by
\begin{align}
    \mathcal{Q}_{xx}=\mathcal{Q}_{yy} &= \frac{1}{4\pi}\begin{pmatrix}
        n & 0 \\
        0 & n+2
    \end{pmatrix}, \label{ap:eq:non-Abelian_QGT_xx}\\
    \mathcal{Q}_{xy}=(\mathcal{Q}_{yx})^{\dagger} &= \frac{i}{4\pi}\begin{pmatrix}
        -n & 0 \\
        0 & n+2
    \end{pmatrix}, \label{ap:eq:non-Abelian_QGT_xy}
\end{align}
where the first column and row represent $n$-th Landau level while the second column and row represent $n+1$-th Landau level.

In particular, for filling factor $\nu=2$, $n=0$ and $n=1$ are filled and thus the non-Abelian quantum geometric tensor is given by setting $n=0$ in Eq.~\eqref{ap:eq:non-Abelian_QGT_xx}, \eqref{ap:eq:non-Abelian_QGT_xy}.

\subsection{Useful formulas for calculations of Landau levels} \label{ap:sec:formulas}
We summarize here useful formulas for the calculation of Landau levels. $\phi_n(x)$ introduced in Eq.~\eqref{ap:eq:varphi} satisfies the following useful properties:
\begin{align}
    &\int_{-\infty}^{\infty}\varphi_n(x)\varphi_m(x)\dd{x}=\delta_{nm} \\
    &\int_{-\infty}^{\infty}\varphi_n(x)\dv{\varphi_m(x)}{x}\dd{x} = \frac{1}{l}\qty(\sqrt{\frac{m}{2}}\delta_{n+1,m}-\sqrt{\frac{n}{2}}\delta_{n,m+1}) \\
    &\int_{-\infty}^{\infty}\varphi_n(x)x\varphi_m(x)\dd{x}=l\qty(\sqrt{\frac{m}{2}}\delta_{n+1,m} + \sqrt{\frac{n}{2}}\delta_{n,m+1}) \\
    &\int_{-\infty}^{\infty}\varphi_n(x)x^2\varphi_m(x)\dd{x}=\frac{l^2}{2}\qty(\delta_{n,m+2}\sqrt{n(n-1)} + (2n+1)\delta_{nm} + \sqrt{m(m-1)}\delta_{n+2,m}) \\
    &\pdv{\varphi_n}{x}=-\frac{x}{l^2}\varphi_n(x) + \frac{\sqrt{2n}}{l}\varphi_{n-1}(x)
\end{align}
With these, we can show the following formulas for the cell-periodic Bloch wavefunctions:
\begin{align}
    \ip{\partial_x u_n}{\partial_x u_n} &= \frac{k_y^2}{e^2 B^2} + \frac{1}{\abs{eB}}(n+1/2) \\
    \ip{\partial_y u_n}{\partial_y u_n} &= \frac{n+1/2}{\abs{eB}} \\
    \ip{\partial_x u_n}{\partial_y u_m} &= \frac{i}{2eB} \\
    \ip{\partial_x u_n}{u_m} = - \ip{u_{n}}{\partial_x u_m} &= i\frac{k_y}{eB}\delta_{nm} + il\qty(\sqrt{\frac{m}{2}}\delta_{n+1, m} + \sqrt{\frac{n}{2}}\delta_{n, m+1}) \\
    \ip{\partial_y u_n}{u_m} = - \ip{u_{n}}{\partial_y u_m} &= \frac{1}{eB l}\qty(\sqrt{\frac{m}{2}}\delta_{n+1,m}-\sqrt{\frac{n}{2}}\delta_{n,m+1})
\end{align}

\end{widetext}

\end{document}